

\magnification=\magstep1

\hsize=37.5truepc
\vsize=58truepc

\baselineskip=12pt
\hfuzz=1pt
\vfuzz=2pt
\tolerance=5000
\parindent=12pt
\voffset=24pt

\newcount\secno      
\newcount\subno      
\newcount\appno      
\newcount\tabno      
\newcount\figno      
\newcount\eqcount    
%
%
%

%

%
\font\eightrm=cmr8
\font\eighti=cmmi8
\font\eightsy=cmsy8
\font\eightbf=cmbx8
\font\eightit=cmti8
\font\eightsl=cmsl8
\font\sixrm=cmr6
\font\sixi=cmmi6
\font\sixsy=cmsy6
\font\sixbf=cmbx6
%
%
\def\tenpoint{\def\rm{\fam0\tenrm}%
  \textfont0=\tenrm \scriptfont0=\sevenrm
                      \scriptscriptfont0=\fiverm
  \textfont1=\teni  \scriptfont1=\seveni
                      \scriptscriptfont1=\fivei
  \textfont2=\tensy \scriptfont2=\sevensy
                      \scriptscriptfont2=\fivesy
  \textfont3=\tenex   \scriptfont3=\tenex
                      \scriptscriptfont3=\tenex
  \textfont\itfam=\tenit  \def\it{\fam\itfam\tenit}%
  \textfont\slfam=\tensl  \def\sl{\fam\slfam\tensl}%
  \textfont\bffam=\tenbf  \scriptfont\bffam=\sevenbf
                            \scriptscriptfont\bffam=\fivebf
                            \def\bf{\fam\bffam\tenbf}%
  \normalbaselineskip=12pt
  \setbox\strutbox=\hbox{\vrule height8.5pt depth3.5pt width0pt}%
  \let\sc=\eightrm \normalbaselines\rm}
%
%
\def\eightpoint{\def\rm{\fam0\eightrm}%
  \textfont0=\eightrm \scriptfont0=\sixrm
                      \scriptscriptfont0=\fiverm
  \textfont1=\eighti  \scriptfont1=\sixi
                      \scriptscriptfont1=\fivei
  \textfont2=\eightsy \scriptfont2=\sixsy
                      \scriptscriptfont2=\fivesy
  \textfont3=\tenex   \scriptfont3=\tenex
                      \scriptscriptfont3=\tenex
  \textfont\itfam=\eightit  \def\it{\fam\itfam\eightit}%
  \textfont\slfam=\eightsl  \def\sl{\fam\slfam\eightsl}%
  \textfont\bffam=\eightbf  \scriptfont\bffam=\sixbf
                            \scriptscriptfont\bffam=\fivebf
                            \def\bf{\fam\bffam\eightbf}%
  \normalbaselineskip=10pt
  \setbox\strutbox=\hbox{\vrule height7pt depth2pt width0pt}%
  \let\sc=\sixrm \normalbaselines\rm}
%
%
\def\title#1
    {\null
    {\pretolerance=10000
     \baselineskip=18pt
     \raggedright
     \noindent
     \tenpoint\bf#1\par}}
\def\author#1
    {\vskip3pc
    {\noindent
     \rm#1\par}
     \smallskip}
\def\address#1
   {{\noindent
     \eightpoint
     \rm#1\par}}
\def\beginabstract
    {\vskip12pt plus3pt minus3pt
     \parindent=18pt
     \eightpoint
     \hang
    {\bf Abstract. }
     \rm }
\def\endabstract
   {\par
    \parindent=12pt
    \tenpoint}
\def\section#1
    {\vskip24pt plus12pt minus12pt
     \bigbreak
     \noindent
     \subno=0
     \global\advance\secno by 1
     \noindent {\bf \the\secno. #1\par}
     \nobreak
     \bigskip
     \noindent}
\def\subsection#1
    {\vskip-\lastskip
     \vskip18pt plus6pt minus6pt
     \bigbreak
     \noindent
     \global\advance\subno by 1
     \noindent {\it \the\secno.\the\subno. #1\par}
     \nobreak
     \medskip
     \noindent}
\def\appendix#1
   {\vskip-\lastskip
    \vskip24pt plus12pt minus12pt
    \subno=0
    \global\advance\appno by 1
    \noindent {\bf Appendix \the\appno. #1\par}
    \nobreak
    \bigskip
    \noindent}
\def\ack
   {\vskip-\lastskip
    \vskip24pt plus 12pt minus12pt
    \bigbreak
    \noindent{\bf Acknowledgments\par}
    \nobreak
    \bigskip
    \noindent}
\def\table#1
   {\midinsert
   {\parindent=18pt
    \hang
    \eightpoint
    \global\advance\tabno by 1
    \bf Table \the\tabno.
    \rm #1\par}
    \medskip
    \eightpoint}

\def\figure#1#2
   {\midinsert
    \vskip#1
    \vskip1pc plus1.5pc
   {\noindent
    \eightpoint
    \global\advance\figno by 1
    \bf Figure \the\figno.
    \rm #2
    \par}
    \vskip 0pt plus2pc
    \endinsert}
\def\shortfigure#1#2
   {\midinsert
    \vskip#1
    \vskip1pc plus1.5pc
   {\centerline
    \eightpoint
    \global\advance\figno by 1
    \bf Figure \the\figno.
    \rm #2
    \par}
    \vskip 0pt plus2pc
    \endinsert}
\def\eq{\global\advance\eqcount by 1
    (\the \eqcount)}
\def\references
    {\vskip24pt plus4pt minus4pt
     \bigbreak
    {\noindent \bf References\par}
     \nobreak
     \eightpoint
     \parindent=0pt
     \bigskip}
\def\ref#1
   {\hangindent=12pt
    \hangafter=1
   {\eightpoint\noindent\frenchspacing
    #1}
    \par}
\def\numref#1#2
   {\parindent=18pt
    \hang \noindent \eightpoint\rm
    \hbox to 15pt{\hss #1\hss}
   {\frenchspacing
    #2}
    \par}
%
\def\e{{\rm e}}       

\def\i{\ifmmode{\rm i}\else\char"10\fi} 
\def\frac#1#2{{#1\over#2}}              


\voffset=-0.15in   

\def\a{\alpha}

\def\x{{\bf x}}
\def\cal{\fam 2}
\def\half{{\frac 12}}

\title{The Interpretation of Quantum Cosmological
Models\footnote{$^{*}$}{\eightpoint To appear in the proceedings of the 13th
Interntational Conference on General Relativity and Gravitation, Cordoba,
Argentina, 1992.} }
\author{Jonathan J. Halliwell}
\address{Center for Theoretical Physics,
Laboratory for Nuclear Science,
Massachusetts Institute of Technology,
Cambridge,
MA 02139, USA.\footnote{$^{\dag}$}{\eightpoint Address after September 1, 1992:
Blackett Laboratory, Imperial College of Science and Technology, South
Kensington, London, SW7 2BZ, UK.} }

\beginabstract
{We consider the problem of extracting physical predictions from the wave
function of the universe in quantum cosmological models.  We state the features
of quantum cosmology an interpretational scheme should confront. We discuss the
Everett interpretation, and extensions of it, and their application to
quantum cosmology. We review the steps that are normally taken in the process
of extracting predictions from solutions to the Wheeler-DeWitt equation for
quantum cosmological models. Some difficulties and their possible resolution
are discussed. We conclude that the usual wave function-based approach admits
at best a rather heuristic interpretation, although it may in the future be
justified by appeal to the decoherent histories approach.}
\endabstract
\bigskip
\centerline{CTP preprint \#2130, July 1992}
\section{Introduction}

Quantum mechanics was originally developed to account for a number of otherwise
unexplained phenomena in the microscopic domain. The scope of the theory
was not thought to extend beyond the microscopic. Indeed, the existence of an
external, macroscopic, classical domain was felt to be necessary for the
theory's interpretation. This view of quantum mechanics has persisted for a
very long time, with not one shred of experimental evidence against it.

Today, however, more ambitious views of quantum mechanics are entertained. The
experimental possibilities afforded by SQUIDS suggest that quantum coherence
may act on macroscopic scales (Leggett, 1980). Furthermore, the separation of
classical and quantum domains is often felt unnatural from a foundational point
of view. The domain of applicability of quantum mechanics may therefore need to
be extended. But in extrapolating quantum mechanics to the macroscopic scale,
where do we stop? At the scale of large molecules? At the laboratory scale? At
the planetary scale? There is only one natural place -- at the scale of the
entire universe. One rapidly arrives, therefore, at the subject of quantum
cosmology, in which quantum mechanics is applied to the entire universe. It
might only be through this subject that the form of quantum mechanics as we
know it may be fully understood.

Yet quantum cosmology was not originally developed with the foundational
problems of quantum mechanics in mind. Rather, it was perhaps viewed as a
natural area to investigate as part of a more general drive to understand
quantum gravity (DeWitt, 1967; Misner, 1969; Wheeler, 1963, 1968). More
recently, the success of models like inflation (Guth, 1981) in early universe
cosmology has led to a greater desire to understand the initial conditions with
which the universe began. On very general grounds, the universe appears to have
emerged from an era in which quantum gravitational effects were important.
Quantum cosmology, in which both matter and gravitational fields are taken to
be quantized, is therefore the natural framework in which to address questions
of initial conditions. Indeed, there has been a considerable amount of
activity in this subject in
recent years.\footnote{$^{\dag}$}{\eightpoint
For general reviews of quantum cosmology, see Fang and Ruffini (1987), Fang and
Wu (1986), Halliwell (1990, 1991), Hartle (1985, 1986, 1990), Hawking (1984b),
Hu (1989), Laflamme (1991), Page (1986, 1991).}

Much of this recent attention on quantum cosmology has been focused on
obtaining a crude idea of the physical predictions made by certain theories of
initial conditions. However, little attention has been devoted to understanding
or even clarifying the principles involved in extracting these predictions from
the mathematical formalism. This contribution represents a small step
towards filling this gap. That is, it is concerned with the {\it
interpretation} of quantum cosmology. In particular, I shall be addressing the
question, ``How are predictions extracted from a given wave function of the
universe?''. I will state immediately that there is no wholly satisfactory
answer to this question. The aim of this contribution, therefore, is to
describe the difficulties involved and the attempts to overcome them, the
procedures that are actually used in practice and the motivation behind them.

The interpretation of quantum cosmology raises two particular sets of issues
which may be discussed separately. The first set concerns the fact that, unlike
ordinary quantum mechanics, the system under scrutiny, the universe, is a
closed and isolated system. The special features of (non-relativistic) quantum
mechanics applied to such systems are discussed in the accompanying paper by
Hartle (Hartle, 1992a). For completeness, some of these features are also
covered here, but from a somewhat different angle (namely from the wave
function point of view, rather than that of the decoherent histories approach).
The second set of issues concerns the fact that the wave function for the
system is described not by a time-dependent Schr\"odinger equation, but by the
Wheeler-DeWitt equation. Associated with this is the so-called ``problem of
time'', and the absence of the usual machinery of projection operators, unitary
evolution, Hilbert space {\it etc.} It is this second set of issues that will
form the primary focus of this paper.

To focus the discussion, consider the sort of features of the universe
one might hope to predict in quantum cosmology. Some of the most important
are:

\item{$\bullet$} {\bf Spacetime is Classical.} One of the crudest and most
obvious observations about the universe we can make, is that the structure of
spacetime is described very accurately by classical laws. The very first
requirement of a quantum theory of the universe, therefore, is that it predict
this. The question of how classical behaviour emerges in quantum systems is a
difficult one, that has not been completely solved even in non-relativistic
quantum mechanics. Still, it ought be possible, at least in some crude sense,
to see whether a given wave function for the universe is consistent with a
prediction of classical spacetime.

\item{$\bullet$} {\bf Initial Inflationary Phase.} Many current observational
features of the universe, such as the horizon, flatness and monopole problems,
are explained by postulating an inflationary phase at very early times. In
models of the universe which admit inflationary solutions, the occurrence and
amount of inflation are generally dependent on initial conditions. A correct
quantum theory of initial conditions should supply the initial conditions for
an inflationary phase to take place.

\item{$\bullet$} {\bf Spectra of Density Fluctuations and Gravitational Waves}.
In inflationary models, it is possible to generate gravitational waves and
density fluctuations consistent with the observed isotropy of the microwave
background, yet sufficiently large for the formation of large scale structure.
These results are obtained by considering the quantum field theory of the
fluctuations during the inflationary phase. Again they are initial conditions
dependent, and one might hope that quantum cosmology will supply the requisite
initial conditions.

\item{$\bullet$} {\bf Low Entropy Initial State.}  One of the most striking
features of the universe is its time asymmetry. It is very far from
equilibrium, indicating that it started out in a very special, low entropy
initial state. One would therefore hope to predict this very smooth beginning.
This is closely related to the spectra of fluctuations.

\noindent In what follows, we will not be concerned with detailed predictions
of particular theories of initial conditions. Rather, we will assume that we
are given a wave function for the universe (perhaps determined by {\it some}
theory of initial conditions), and discuss the extraction of predictions from
it. In particular, we will focus on the emergence of a {\it semiclassical
domain} -- an approximately classical spacetime background with quantum fields
in it\footnote{$^{\dag}$}{\eightpoint This is, loosely speaking, the same
thing as the {\it quasi-classical} domain defined by Gell-Mann and Hartle
(1990) in the context of the decoherent histories approach, but I give it a
slightly different name since the wave function approach considered here
does not permit it to be defined in quite the same way.}. This is
appropriate for two reasons. First, all observational features of the universe
are semiclassical in nature. Second, even if the theory could make observable
quantum gravitational predictions, as one would certainly hope of a quantum
theory of cosmology, observation of them would be through their correlation
with the semiclassical domain. Quite generally, therefore, the emergence of the
semiclassical domain is the appropriate thing to look for.

\section{Canonical Quantization}

In the interests of conciseness, I shall assume that the formalism of quantum
cosmology is known, and give only the briefest account here (see for example
Halliwell (1990), and the general
references cited earlier). For definiteness, we consider the canonical
quantization of Einstein gravity coupled to matter for closed universes. The
quantum state of the system, the universe, is represented by a wave
functional, $\Psi[h_{ij}, \phi]$, a functional on superspace, the space of
three-metrics $h_{ij}$ and matter fields $\phi$ on a three-surface. The wave
function has no explicit dependence on time. There is therefore no
Schr\"odinger equation, only the constraints of the Dirac quantization
procedure,
$$
\hat {\cal H} \Psi[h_{ij}, \phi]= 0, \quad \quad \quad \hat {\cal
H}_i \Psi [h_{ij}, \phi]= 0
\eqno(2.1)
$$
where $\hat {\cal H}$ and $\hat {\cal H}_i$ are the operator versions of the
classical constraints,
$$
\eqalignno{
{\cal H} &= h^{-\half} \left(\pi^{ij} \pi_{ij} - \half \pi^i_i \pi^j_j \right)
- h^{\half} \left( \ ^3R- 2 \Lambda \right) + {\cal H}^{matter} = 0 &(2.2) \cr
{\cal H}_i &= -2 \pi_{ij}^{\ |j} + {\cal H}_i^{matter} = 0 &(2.3) \cr }
$$
(in the usual notation). Eqs.(2.1) are of course the Wheeler-DeWitt equation
and the momentum constraints.

Wave functions satisfying the constraints (2.1) may also be generated using a
(complex) path integral representation (Halliwell and Hartle, 1990, 1991). This
is used, for example, in the specification of the no-boundary wave function
(Hawking, 1982, 1984a; Hartle and Hawking, 1983). The sum-over-histories
approach is more general in that it may be used to construct more complicated
amplitudes ({\it e.g.}, ones depending on the three-metric and matter fields on
many three-surfaces) and in fact the latter type of amplitude is likely to be
the most useful for interpreting the theory. Here we are concerned with
describing what has actually been done, and this means discussing single
surface amplitudes.

There are other very different ways of quantizing the classical system
described by the constraints, (2.2), (2.3). In particular, one could
contemplate solving the constraints classically, prior to quantization, and
then quantizing an unconstrained theory. The Dirac quantization scheme outlined
here is the one most commonly used in practical quantum cosmology, and it is
this that we shall use in what follows.

Of course, all known approaches to quantum gravity are fraught with severe
technical difficulties, and the formal framework outlined above is no
exception. Moreover, a proper theory of quantum gravity, should it exist, might
involve substantial departures from this general framework ({\it e.g.}, string
theory). However, there are a number of reasons why it might nevertheless make
sense to persevere with this approach to quantum cosmology. Perhaps the most
important is that in quantum cosmology one is frequently concerned with {\it
issues}. Many issues, and in particular the interpretational issues considered
here, are not very sensitive to the resolution of technical difficulties.
Furthermore, they are likely to be equally present in any approach to
quantum gravity. No generality is lost, therefore, in employing the formal
scheme envisaged in Eq.(2.1).

\section{Interpretation}

The question of interpretation is that of extracting physical statements about
the universe from the wave function, $\Psi[h_{ij}, \phi]$. In non-relativistic
quantum mechanics there is a well-defined procedure for achieving this. It is
generally known as the Copenhagen interpretation. Although it is frequently
regarded as problematic from a foundational point of view, it has been very
successful in its physical predictions.\footnote{$^{\dag}$}{\eightpoint
Many of the key papers on the Copenhagen interpretation may be found in Wheeler
and Zurek (1983).} The Copenhagen interpretation involves a number of features
and assumptions that should be highlighted for the purposes of this account:

\item{\bf C1.} It assumes the existence of an {\it a priori} split of the world
into two parts: a (usually microscopic) quantum system, and an external
(usually macroscopic) classical agency.

\item{\bf C2.} It concerns systems that are not genuinely closed, because they
are occasionally subject to intervention by the external agency.

\item{\bf C3.} The process of prediction places heavy emphasis on the notion of
{\it measurement} by the external agency.

\item{\bf C4.} Predictions are probabilistic in nature, and generally only have
meaning when measurements are performed on either a large ensemble of identical
systems, or on the same system many times, each time prepared in the same
state.

\item{\bf C5.} Time plays a very distinguished and central role.

\noindent Quantum cosmology, by contrast, involves a number of corresponding
features and assumptions that render the Copenhagen interpretation woefully
inadequate:

\item{\bf QC1.} It is assumed that quantum mechanics is universal, applying to
microscopic and macroscopic systems alike, up to and including the entire
universe. There can therefore be no {\it a priori} split of the universe into
quantum and classical parts.

\item{\bf QC2.} The system under scrutiny is the entire universe. It is a
genuinely closed and isolated system without exterior.

\item{\bf QC3.} Measurements cannot play a fundamental role, because there can
be no external measuring apparatus. Even internal measuring apparatus should
not play a role, because the conditions in the early universe were so extreme
that they could not exist.

\item{\bf QC4.} The universe is a unique entity. It does not belong to an
ensemble of identical systems; nor is it possible to make repeated measurements
on it prepared in the same state.

\item{\bf QC5.} The problem of time: general relativity does not obviously
supply the time parameter so central to the formulation and interpretation of
quantum theory.

\noindent Of these features, only (QC5) is specific to quantum gravity. The
rest would be true of any closed and isolated system described by
non-relativistic quantum mechanics, and may be discussed in this context.

The problem of interpretation is that of finding a scheme which confronts
features (QC1)--(QC5), yet which may be shown to be consistent with or to
reduce to (C1)--(C5) under suitable approximations and restrictions.

Probably the best currently available approach to these difficulties is the
decoherent histories approach, which employs not the wave function, but the
decoherence functional as its central tool.\footnote{$^{\dag}$}{\eightpoint The
consistent histories approach to quantum mechanics was introduced by Griffiths
(1984), and later developed by Omn\`es (reviewed in Omn\'es, 1990, 1992). Much
of it was developed independently under the name decoherent histories, by
Gell-Mann and Hartle (1990). See also Hartle (1990, 1992a). For applications
and generalizations, see Blencowe (1991), Albrecht (1990) and Dowker and
Halliwell (1992).} It has a number of features which strongly recommend it for
quantum cosmology: it specifically applies to closed systems; it assumes no
{\it a priori} separation of classical and quantum domains; it does not rely on
the notion of measurement or observation; and its focus on histories rather
than events at a single moment of time might sidestep (or at least alleviate)
the problem of time in quantum gravity. Unfortunately, a detailed application
of this approach to quantum cosmology has not yet been carried out (although
efforts in this direction are currently being made (Hartle, 1992b)).
Furthermore, my task here is to describe what has actually been done, and for
that reason I will describe the somewhat cruder approaches to interpretation
based on the wave function. However, I find it useful to remain close to the
spirit of the decoherent histories approach, and to think of the wave function
approach as an approximation to it (although in a sense yet to be explained).
In particular, it is convenient to have as one's aim the assignment of
probabilities to histories of the universe.

In order to deal with the issues noted above, conventional wave function-based
approaches invoke the Everett (or ``Many Worlds'')
interpretation.\footnote{$^{\dag\dag}$}{\eightpoint Everett (1957). A useful
collection of papers on the Everett interpretation may be found in DeWitt and
Graham (1973). See also Bell (1981), Deutsch (1985), Kent (1990), Smolin (1984)
and Tipler (1986).} Above all, the Everett interpretation is a scheme
specifically designed for quantum mechanical systems that are closed and
isolated. Everett asserted that quantum mechanics should be applicable to the
entire universe, and there should be no separation into quantum and classical
domains. These features of the Everett interpretation are therefore consistent
with features (QC1) and (QC2) of quantum cosmology. Furthermore, the state of
the entire system should evolve solely according to a wave equation, such as
the Schr\"odinger equation, or in quantum cosmology the Wheeler-DeWitt
equation, and there should be no discontinuous changes (collapse of the wave
function).

Everett went on to model the measurement process by considering a world divided
into a large number of subsystems, and showed how the conventional Copenhagen
view of quantum mechanics can emerge. It is this part of the Everett
interpretation that leads to its ``Many Worlds'' feature -- the idea that the
universe splits into many copies of itself whenever a measurement is performed.
It is at this stage, however, that the original version of the Everett
interpretation departs in its usefulness from practical quantum cosmology. For
the sort of models one is often interested in interpreting in quantum cosmology
are minisuperspace models, which are typically very simple, and do not contain
a large number of subsystems. Furthermore, the many worlds aspect of the
interpretation has, I believe, been rather over-emphasized, perhaps at the
expense of undermining the credibility of the overall set of ideas. The Everett
interpretation, especially as it applies to practical quantum cosmology, is not
so much about many worlds, but rather, about how one might make sense of
quantum mechanics applied to genuinely closed systems.

It is, therefore, convenient to pass to a restatement of the Everett
interpretation due to Geroch (1984). Geroch translated Everett's modeling of
the measurement process using a large number of subsystems into statements
about the wave function of the entire system. He argued that predictions for
closed quantum systems essentially boil down to statements about ``precluded''
regions -- regions of configuration space in which the wave function for the
entire system is very small. From here, it is a small step to a slightly more
comprehensive statement given by Hartle, specifically for quantum cosmology
(Hartle, 1986). It is the following:

\item{} {\it If the wave function for the closed system is strongly peaked
about a particular region of configuration space, then we predict the
correlations associated with that region; if it is very small, we predict the
lack of the corresponding correlations; if it is neither strongly peaked, nor
very small, we make no prediction}.

\noindent This basic idea appears to have been adopted,
perhaps provisionally, in most, if not all, attempts to interpret the wave
function in quantum cosmology (see also Wada (1988)).

The statement is admittedly rather vague and a number of qualifying remarks are
in order. Firstly, to say that a wave function is ``strongly peaked''
necessarily involves some notion of a measure, and this has to be specified.
For example, do we use $ | \Psi|^2 $? The Klein-Gordon current constructed from
$\Psi$? We will return to this below. Second, the interesting correlations in
the wave function are often not evident in the configuration space form. It is
therefore appropriate to interpret the above statement rather broadly, as
refering to not just the wave function, but any distribution constructed from
the wave function. For example, phase space distributions such as the Wigner
function, and related functions, have been the focus of attention in a number
of papers, and these have sometimes proved quite useful in identifying
classical correlations (Anderson, 1990; Calzetta and Hu, 1989; Habib, 1990;
Habib and Laflamme, 1990; Halliwell, 1987, 1992; Kodama, 1988; Singh and
Padmanabhan, 1989).

For practical quantum cosmology, the implication of the above interpretational
scheme is that rather than find the probability distributions for quantities
felt to be of interest, as in ordinary quantum mechanics, it is necessary to
determine those quantities for which the theory gives probabilities close to
zero or one, and hence, for which it makes predictions. On the face of it, this
may seem to suggest that the predictive power of quantum cosmology is very
limited in comparison to ordinary quantum mechanics. However, by studying
isolated systems consisting of a large number of identical subsystems, it may
be shown that this interpretation implies the usual statistical interpretation
of ordinary quantum mechanics, in which subsystem probabilities are given by
$|\psi|^2$  where $\psi$ is the subsystem wave function (Farhi, Goldstone and
Gutmann, 1989; Finkelstein, 1963; Graham, 1973; Hartle, 1968). It is in this
sense that feature (QC4) of quantum cosmology is reconciled with (C4) of the
Copenhagen interpretation.

Given a measure on a set of possible histories for the universe, it is often
not peaked about a particular history or family of histories. To obtain
probabilities close to one or zero, it is often necessary to restrict attention
to a certain subset of the possible histories of the universe, and make
predictions within that subset. That is, one looks at {\it conditional
probabilities}. The motivation behind this is anthropic reasoning -- we as
observers do not look out into a generic universe, but to one in which the
conditions for our own existence have necessarily been realized (see, for
example, Barrow and Tipler, 1986). Obviously the detailed conditions for our
existence could be very complicated and difficult to work out. However, it is
possible to get away with exploiting only the weakest of anthropic assumptions
in order to make useful and interesting predictions about the universe. It is,
for example, extremely plausible that the existence of life requires the
existence of stars like our sun. It therefore makes sense to restrict attention
only to those histories of the universe which exist long enough for stars to
form before recollapsing.

This completes our brief survey of the special features of closed quantum
systems. We have discussed features (QC1), (QC2) and (QC4) of quantum
cosmology. The status of measurements, (QC3), in the above discussion is
admittedly somewhat vague, although it is clear that they do not play a
significant role. Their status is perhaps clarified more fully in the
decoherent histories approach. Finally, although it is an important issue for
the interpretation of quantum cosmology, I have not discussed the problem of
time, (QC5). This will be briefly mentioned below, but it would take a lot of
space to do it justice. The interested reader is referred to the reviews by
Kucha{\v r} (1989, 1992), which cover the problem of time and its connections
with the interpretation of quantum cosmology. See also Unruh and Wald (1989).

\section{Interpretation of Solutions to the Wheeler-DeWitt Equation}

We now come to the central part of this contribution, which is to describe how
in practice predictions are actually extracted from solutions to the
Wheeler-DeWitt equation. As mentioned earlier, the primary aim is to determine
the location and features of the semiclassical domain. Different papers in the
literature treat the process of prediction in different ways, but it seems to
boil down to four distinct steps, which I now describe in turn.

\vskip 0.15in
\noindent {\bf A. Restriction to Perturbative Minisuperspace.}
\vskip 0.05in

\noindent The full theory described by Eq.(2.1) is not only difficult to handle
technically, it is not even properly defined. This problem is normally avoided
by artificially restricting the fields to lie in the region of superspace in
the neighbourhood of homogeneity (and often isotropy). That is, one restricts
attention to the finite dimensional subspace of superspace called
``minisuperspace'', and considers small but completely general inhomogeneous
perturbations about it.\footnote{$^{\dag}$}{\eightpoint This general approach
has been the topic of many papers, including Banks (1985), Banks,
Fischler and Susskind (1985), Fischler, Ratra and Susskind (1986), Halliwell
and Hawking (1985), Lapchinsky and Rubakov (1979), Shirai and Wada (1988),
Vachaspati and Vilenkin (1988), Wada (1986).}

In more detail, the sort of restrictions entailed are as follows. One restricts
the three-metric and matter fields to be of the form,
$$
h_{ij}(\x,t) = h^{(0)}_{ij} (t) + \delta h_{ij} (\x,t ), \quad \Phi(\x, t) =
\phi(t) + \delta \phi (\x, t)
\eqno(4.1)
$$
Here, the minisuperspace background
is described by the homogeneous fields $ h^{(0)}_{ij} $ and $\phi(t)$. For
example, the three-metric could be restricted to be homogeneous and isotropic,
described by a single scale factor $a$. We will denote the minisuperspace
background coordinates by the finite set of functions $q^{\a}(t)$, where
$\a=1,\cdots n $. $\delta h_{ij}$ and $\delta \phi $ are small inhomogeneous
perturbations about the minisuperspace background, describing gravitational
waves and scalar field density perturbations. They are retained only up to
second order in the action and Hamiltonian (and therefore to first order in the
field equations). For convenience, we will denote the perturbation modes simply
by $\delta \phi$.

With these restrictions on the class of fields considered, the Wheeler-DeWitt
equation, after integration over the three-surface, takes the form $$ \left[ -
{1 \over 2 m_p^2 } \nabla^2 + m_p^2 U(q) + H_2 ( q, \delta \phi ) \right] \Psi
(q, \delta \phi ) = 0 \eqno(4.2) $$ Here, $ \nabla^2 $ is the Laplacian
operator in the minisuperspace modes $q$ and we have explicitly included the
Planck mass $m_p$, since it is to be used as a large parameter in a
perturbative
expansion. $H_2$ is the Hamiltonian of the perturbation modes, $\delta \phi$,
and is quadratic in them. There are more constraint equations associated with
the remaining parts of the Wheeler-DeWitt equation, and with the momentum
constraints. These are all linear in the perturbations and can be solved,
after gauge-fixing. When this is done, only Eq.(4.2) remains, in which $\delta
\phi$ may be thought of as a gauge-invariant perturbation variable.

The restriction to perturbative minisuperspace is of course very difficult to
justify from the point of view of the full theory. A number of attempts have
been made to understand the sense in which minisuperspace models might be part
of a systematic approximation to the full theory, but the answer seems to be
that their connection is at best tenuous (Kucha{\v r} and Ryan, 1986, 1989).
What one can say, however, is that solutions to the minisuperspace field
equations, including perturbations about them, will (with a little care) be
solutions to the full field equations, and thus the lowest order semiclassical
approximation to perturbative minisuperspace quantum cosmological models will
agree with the lowest order semiclassical approximation to the full theory.
These models may therefore be thought of as useful models in which a number of
issues can be profitably investigated, but which may also give some crude
predictions about the physical universe.

The essential ideas of the practical interpretational scheme described here
will very probably be applicable to situations more general than perturbative
minisuperspace, but very little work on such situations has been carried out.
We will not go into that here.

\vskip 0.15in
\noindent {\bf B. Identification of the Semiclassical Regime.}
\vskip 0.05in

\noindent The next step involves inspecting the wave function $\Psi(q^{\a},
\delta \phi)$, asking how it behaves as a function of the minisuperspace
variables $q$, and in particular, identifying the regions in which the wave
function is exponentially growing or decaying in $q$, or oscillatory in $q$.

The regions in which the wave function is rapidly oscillating in $q$ are
regarded as the semiclassical domain, in which the modes $q$, are approximately
classical, whilst the perturbation modes $\delta \phi$ need not be. This
interpretation comes partly from analogy with ordinary quantum mechanics. But
also one can often argue that certain distributions constructed from a rapidly
oscillating wave function are peaked about classical configurations. For
example, the Wigner function, mentioned earlier, is often used to support this
interpretation.

The other regions, in which the wave function tends to be predominantly
exponential in behaviour are regarded as non-classical, like the
under-the-barrier wave function in tunneling situations. Were this ordinary
quantum mechanics, then the wave function would typically be exponentially
small in the tunneling regions. One could then invoke Geroch's version of the
Everett interpretation, and just that the system will not be found in this
region, because it is ``precluded''. However, a peculiar feature of gravity
(readily traced back to the indefiniteness of the action) is that the wave
function may be either exponentially small or exponentially large in the
regions where it is of exponential form. Nevertheless, one still says that the
system is not approximately classical in this regime, even when the wave
function is exponentially large. To support this claim, one can argue that, in
contrast to the oscillatory regions, a predominantly exponential wave function,
either growing or decaying, is {\it not peaked} about classical configurations.

\vskip 0.15in \noindent {\bf C. WKB Solution in the Oscillatory Regime.} \vskip
0.05in

\noindent The next stage of the scheme involves solving in more detail in the
region in which the wave function is rapidly oscillating in the minisuperspace
variables. This involves focusing on a particular type of state, namely the WKB
state, $$ \Psi(q, \delta \phi)= C (q) \ \exp\left(im_p^2 S_0(q) \right)\
\psi(q,\delta \phi) + O(m_p^{-2}) \eqno(4.3) $$ $S_0(q)$ is real, but $\psi$
may be complex. Many models also involve a slowly varying exponential prefactor
contributing at order $m_p^2$, but for the purposes of the present discussion
it can be assumed that this is absorbed into $C$. Also, a possible phase at
order $m_p^0$ depending only on $q$ may be absorbed into $\psi$, so $C$ may be
taken to be real.

Eq.(4.3), it must be stressed, is an {\it ansatz} for the solution, and many of
the predictions subsequently derived depend on assuming this particular form.
We will return later to the question of the validity and usefulness of this
particular ansatz.

The Wheeler-DeWitt equation may now be solved perturbatively, by inserting the
ansatz (4.3), and using the Planck mass $m_p$ as a large parameter. Since this
parameter is not dimensionless, the expansion is meaningful only on length
scales much greater than the Planck length. Note also that a double expansion
of the full Wheeler-DeWitt equation is involved: a WKB expansion in the Planck
mass, and a perturbation expansion in small inhomogeneities about
minisuperspace.

Equating powers of the Planck mass, one obtains the following. At lowest order,
one gets the Hamilton-Jacobi equation for $ S_0 $, $$ {1\over 2} (\nabla
S_0)^2+U(q)=0. \eqno(4.4) $$ To next order one obtains a conservation equation
for $C$ $$ 2 \nabla S_0 \cdot \nabla C - C \nabla^2 S_0 = 0 \eqno(4.5) $$ and a
Schr\"odinger equation for $\psi$, $$ i \nabla S_0 \cdot \nabla \psi = H_2 \psi
\eqno(4.6) $$

Consider now (4.4). As indicated above, it may be argued that a wave function
predominantly of the form ${\e}^{im_p^2 S_0}$ indicates a strong correlation
between coordinates and momenta of the form $$ p_{\a} = m_p^2 { \partial S_0
\over \partial q^{\a} } \eqno(4.7) $$ Since $ \dot q^{\a} = p^{\a}$, (4.7) is a
set of $n$ first order differential equations (where, recall, $\a=1,\cdots n$).
Furthermore, since by (4.4), $S_0$ is a solution to the Hamilton Jacobi
equation, it may be shown that these equations define a first integral to the
classical field equations. The first integral (4.7) may be solved to yield an
$n$-parameter set of classical solutions. It is for this reason that one says
that the wave function (4.3) to leading order, corresponds to an ensemble of
classical solutions to the field equations.

In ordinary quantum mechanics, this interpretation may be substantiated by
subjecting an initial wave function of the form (4.3) to a sequence of
approximate position samplings, and showing that the resulting probability
distribution for histories is peaked about the set of classical solutions
satisfying the first integral (4.7). See Halliwell and Dowker (1992), for
example, for efforts in this direction.

The tangent vector to this congruence of classical paths is $$ \nabla S_0 \cdot
\nabla \ \equiv \ {\partial \over \partial t} \eqno(4.8) $$ (4.6) is therefore
the functional Schr\"odinger equation for the perturbation modes along the
family of minisuperspace trajectories. This indicates that the perturbation
modes are described by quantum field theory (in the functional Schr\"odinger
picture) along the family of classical backgrounds described by (4.7).

\vfill\eject
\vskip 0.15in \noindent {\bf D. The Measure on the Set of Classical
Trajectories} \vskip 0.05in

\noindent The final stage is to put a measure on the congruence of classical
paths, and to find an inner product for solutions to the Schr\"odinger equation
(4.8). Suppose one chooses some $(n-1)$-dimensional surface in minisuperspace
as the beginning of classical evolution. Through (4.7), the wave function then
effectively fixes the initial velocities on that surface, in terms of the
coordinates. However, the wave function should also provide a probability
measure on the set of classical trajectories about which the wave function is
peaked. How is this measure constructed from the wave function? The
construction of a satisfactory non-negative measure remains an outstanding
problem of quantum cosmology, but perhaps the most successful attempts so far
involve the Klein-Gordon current,
$$
J = {i \over 2} \left( \Psi \nabla \Psi^*
- \Psi^* \nabla \Psi \right)
\eqno(4.9)
$$
It is conserved by virtue of the
Wheeler-DeWitt equation (4.2), $$ \nabla \cdot J = 0 \eqno(4.10) $$ Choose a
family of surfaces $\{ \Sigma_{\lambda} \}$, parametrized by $\lambda$, cutting
across the flow of $J$. Then (4.10) suggests that for each $\lambda$, a
probability measure on the congruence of trajectories is the flux of $J$ across
the surface: $$ dP = J \cdot  d \Sigma \eqno(4.11) $$ This measure is conserved
along the flow of $J$, as is readily shown from (4.10).

The problem with (4.11), however, is that it is not always positive. For
example, if the surfaces $\Sigma$ are taken to be surfaces of constant scale
factor, the flow of $J$ typically cuts these surfaces more than once, because
of the possibility of expanding and collapsing universes, leading to negative
values for (4.11). Furthermore, $J$ vanishes when $\Psi$ is real. Still, some
sense may be made out of (4.11) by restricting to the WKB wave functions,
(4.3). For these wave functions, the current is $$ J \approx m_p^2 \ | C|^2 \ |
\psi|^2 \ \nabla S_0 + O (m_p^0) \eqno(4.12) $$ For reasonably large regions of
minisuperspace (but not globally), it is usually possible to choose a set of
surfaces $\{ \Sigma_{\lambda} \} $ for which $ \nabla S_0\cdot d \Sigma \ge 0
$, and thus the probability measure will be positive. Furthermore, this measure
implies the standard $|\psi|^2$ measure for the perturbation wave functions,
completing the demonstration that quantum field theory for the perturbations
emerges in the semiclassical limit.

This approach was described a long time ago by Misner (1972), and developed
more recently by Vilenkin (1989). It is problematic for a number of reasons:
one is the global problem of choosing the surfaces $\{\Sigma_{\lambda}\}$;
another is that is very tied to the WKB wave functions (4.3). More will be said
about this in the next section. Still, it allows predictions to be made in a
number of situations of interest.

The ``naive'' Schr\"odinger measure is also sometimes proposed in place of
(4.11) ({\it e.g.}, Hawking and Page, 1986, 1988). This is the assertion that
the probability of finding the system in a region of superspace of volume $dV$
is $| \Psi |^2 dV $, where $\Psi$ is the wave function of the whole system.
This does at least have the advantage that is is everywhere positive.
Furthermore, it can be argued to reduce to (4.11) for WKB wave functions, in
the limit in which the volume of superspace $dV$ is taken to be hypersurface of
codimension one slightly thickened along the direction of the flow of $J$. As
argued by Kucha{\v r}, however, this measure is problematic for other reasons
(Kucha{\v r}, 1992).

At this stage, it is appropriate to comment on the problem
of time. In the Wheeler-DeWitt equation (4.2) (or (2.1)), there is no
distinguished variable that plays the role of time. This is
the problem of time. In the scheme described above, however, a time parameter
has emerged. It is the parameter $\lambda$ labeling the family
of surfaces $\{\Sigma_{\lambda} \}$, and may be chosen to be the same as the
parameter $t$ defined in Eq.(4.8), the affine parameter along the integral
curves of $\nabla S_0$. The point to be made is that this parameter has emerged
only in the region where the wave function is oscillatory, and in particular,
as a consequence of the assumed semiclassical form of the wave function, (4.3).
This is therefore an explicit illustration of the point of view, not uncommonly
expressed, that time, and indeed spacetime, need not exist at the most
fundamental level but may emerge as approximate features under some suitable
set of conditions. It is in this sense that feature (QC5) of quantum cosmology
may be reconciled with (C5) of ordinary quantum mechanics.

Modulo the above difficulties, Eq.(4.11) is the desired probability measure on
possible histories of the universe. It is commonly not normalizable over the
entire surface $\Sigma$; but this need not matter, because it is {\it
conditional} probabilities that one is typically interested in. Suppose, for
example, one is given that the history of the universe passed through a subset
$s_1$ of a surface $\Sigma$, and one wants the probability that the universe
passed through a subset $s_0$ of $s_1$. The relevant conditional probability
is, $$ p(s_0|s_1) = { \int_{s_0} J \cdot d \Sigma \over \int_{s_1} J \cdot d
\Sigma } \eqno(4.13) $$ The set of all histories intersecting $\Sigma$ could
include universes which recollapse after a very short time. A reasonable choice
for $s_1$, therefore, might be universes that exist long enough for stars to
form before recollapsing, as discussed earlier. $s_0$ could then be taken to be
the subset of such universes which possess certain features resembling our
universe. If the resulting conditional probability turned out to be close to
one or zero, this would then constitute a definite prediction.

Steps (A)--(D) above constitute the general interpretational scheme implicit or
explicit in most attempts to extract predictions from the wave function of the
universe. It is not by any means a consistent interpretational scheme, but is
almost a list of rules of thumb inspired by the Everett interpretation, and
built on analogies with ordinary quantum mechanics. It has many difficulties,
some of which we now discuss.

\section{Problems and Objections}

We have argued that the WKB wave functions (4.3) correspond to an ensemble of
classical paths defined by the first integral (4.7). Strictly speaking, what
this means is that the WKB wave function is really some kind of {\it
superposition} of wave functions, each of which corresponds to an individual
classical history (like a superposition of coherent
states).\footnote{$^{\dag}$}{\eightpoint For the explicit construction of wave
packets in quantum cosmology, see Kazama and Nakayama (1985) and Kiefer
(1988).} A closely related point is the question of why one should be allowed
to study an interpretation based on the WKB form, (4.3): one would expect a
more general wave function to be expressed as a {\it sum} of such terms. In
each of these cases, we are acting as if we had a classical statistical
ensemble, when what we really have is a superposition of interfering states.
Why should it be permissible to treat each term in the sum separately, when
strictly they are interfering?

This point concerns the general question of why or when it is permissable to
ignore the interference terms in a superposition, and treat each term as if it
were the member of a statistical ensemble. Technically, the destruction of
interference is generally referred to as decoherence.

Two notions of decoherence have been employed. The most precise is that of the
decoherent histories approach, where it enters at a very fundamental level:
Interference is most generally and properly thought of as the failure of the
probability sum rules for quantum-mechanical histories. Decoherence, as
destruction of interference, is therefore best regarded as the recovery of
these rules (Gell-Mann and Hartle, 1990).

By contrast, in the wave function approach to quantum cosmology, decoherence
appears to have been added as an afterthought, using a different and somewhat
vaguer definition: decoherence is held to be related to the tendancy of the
density matrix towards diagonality (Joos and Zeh, 1985). It is also associated
with the establishment of correlations of a given system with its environment,
and with the stability of certain states under evolution in the presence of an
environment (Zurek, 1981, 1982; Unruh and Zurek, 1989). These definitions are
problematic for a number of reasons. One is that (in ordinary quantum
mechanics) the density matrix refers only to a single moment of time, yet the
proper definition of interference -- the effect one is trying to destroy -- is
in terms of histories. Another is the question of the basis in which the
density matrix should be diagonal.\footnote{$^{\dag}$}{\eightpoint The basis
issue is discussed, for example, in Barvinsky and Kamenshchik (1990),
Deutsch (1985), and Markov and Mukhanov (1988). Also, see Zurek (1992) for a
possible reconciliation of the above two differing views of decoherence.}

Despite the difficulties, the density matrix approach has been the topic of a
number of papers on decoherence in quantum cosmology, perhaps because it is
technically much simpler (Fukuyama and Morikawa, 1989; Habib and Laflamme,
1990; Halliwell, 1989; Kiefer, 1987; Laflamme and Louko, 1991; Mellor, 1989;
Morikawa, 1989; Padmanabhan, 1989; Paz, 1991; Paz and Sinha, 1992). Models are
considered in which the variables of interest are coupled to a wider
environment, and a coarse-graining is carried out, in which the states of the
environment are traced over. In the case of the whole universe, which strictly
has no environment, this is achieved quite simply by postulating a sufficiently
complex universe with a suitably large number of subsystems, and ignoring
some of the subsystems. The typical result of such models is that the
interference terms are suppressed very effectively. Furthermore, one and the
same mechanism of coarse-graining also causes decoherence in the
histories-based definition of the process. It is therefore plausible that a
more sophisticated analysis using the decoherent histories approach will lead
to the same conclusions. One way or another, one finds some amount of
justification for treating the terms in a superposition separately, and
treating the set of paths to which a WKB wave function correspond as a
statistical ensemble. These arguments have not gained universal acceptance,
however (see for example, Kucha{\v r} (1992)).

After the presentation of this contribution, J.Ehlers asked about the
observational status of quantum cosmology. Since quantum cosmology aspires to
make observable predictions, this is obviously a very important question.
Interest in quantum cosmology arose partly as a result of the realization that
conventional classical cosmological scenarios relied on certain (possibly
tacit) assumptions about initial conditions. As is well known, the inflationary
universe scenario alleviates the hot big bang model of extreme dependence on
initial conditions; but it does not release it from {\it all} dependence. The
amount of inflation, the details of the density fluctuations generated, and
indeed, the very occurrence of inflation are initial conditions-dependent. One
of the main successes of quantum cosmology (modulo the objections described in
this paper) has been to demonstrate that the desired initial conditions for
inflation are implied by certain simple boundary condition proposals for the
wave function of the universe. This might therefore be regarded as an
observational prediction, although it is admittedly a very indirect one.

More direct tests will clearly be very difficult. The universe has gone through
many stages of evolution, each of which is modeled separately. In observing the
universe today, it is difficult to distinguish between effects largely due to
initial conditions and those largely due to dynamical evolution or to the
modeling of a particular stage. What is needed is an effect produced very early
in the history of the universe, but that is largely insensitive to subsequent
evolution. Grishchuk (1987) has argued that gravitational waves might be the
sought-after probe of the very early universe. Boundary condition proposals for
the wave function of the universe typically make definite predictions about the
initial state of the graviton field ({\it e.g.},  Halliwell and Hawking, 1985).
Memory of this initial state could well be preserved throughout the subsequent
evolution of the universe, because gravitational waves interact so weakly.
Parts of their spectra observable today might therefore contain signatures of
the initial state, leading to the exciting possibility of distinguishing
observationally between different boundary condition proposals. These ideas are
of course rather speculative, and gravitational wave astronomy is still in a
very primitive state. Still, quantum cosmology suffers from an acute lack of
connections observational cosmology, and any potentially observable effect
deserves further study.

L.Smolin commented that the scheme described here is very semiclassical in
nature, and suggested that it is perhaps not much more than a classical
statistical theory. It is certainly true that it is very semiclassical in
nature, that its predictive output has the form of classical statistical
theory, and perhaps causes one to wonder how much of it is really
quantum-mechanical in nature. However, there {\it are} some genuinely
quantum-mechanical aspects to this predictive scheme. Perhaps the principle one
is the prediction of regions in which classical laws are not valid (the regions
of superspace in which the wave function is exponential). Determination of the
existence and location of these regions requires the quantum theory -- their
existence and location cannot be anticipated by inspection of the classical
theory alone. Furthermore, the existence of these regions underscores the
necessity of discussing the issue of initial or boundary conditions from within
the context of the quantum theory. For in classical theories of initial
conditions (including classical statistical ones), one might be attempting to
impose classical initial conditions in a region in which classical laws are
quite simply not valid.

Finally, a critical appraisal of the quantum cosmology program (which
inspired some of the remarks made here) may be found in Isham (1991).

\section{Conclusions}

In this talk I have described the heuristic set of rules that have been used so
far to make crude but plausible predictions in quantum cosmology. These rules
are, however, rather heuristic and semiclassical in nature. The interpretation
of the wave function seems to proceed on a case by case basis, and no
satisfactory scheme for a completely general wave function is available. I am
therefore forced to conclude that quantum cosmology does not yet possess an
entirely satisfactory scheme for the extraction of predictions from
the wave function.

At the present time, the decoherent histories approach appears to offer the
most promising hope of improving the situation. A reasonable expectation is
that the heuristic interpretational techniques described here will emerge from
this more sophisticated approach. Detailed demonstration of this assertion is
very much a matter of current research.

\ack

I thank Jim Hartle and Raymond Laflamme for very useful discussions on the
material of this talk. I am very grateful to the organizers of GR13, and in
particular to Carlos Kozameh and Bob Wald, for inviting me to take part in this
stimulating meeting. I would also like to thank J\"urgen Ehlers for initiating
the session on Interpretational Issues in Quantum Cosmology, and for his
thought-provoking questions. Finally, I am very grateful to Mario Castagnino,
and his colleagues in Buenos Aires and Rosario, for their warm hospitality
during my stay in Argentina.

\references

\def\pr{{\sl Phys.Rev.\ }}  \def\prep{{\sl
Phys.Rep.\ }}  \def\rmp{{\sl Rev.Mod.Phys.\ }}
 \def\np{{\sl Nucl.Phys.\ }} \def\pl{{\sl
Phys.Lett.\ }} \def\annp{{\sl Ann.Phys.(N.Y.)\ }} \def\cqg{{\sl Class.Quantum
Grav.\ }}  
\def\ijmp{{\sl Intern.J.Mod.Phys.\ }} 
\def\ijtp{{\sl Int.J.Theor.Phys.\ }} 
\def\amjp{{\sl Am.J.Phys.\ }} \def\jsp{{\sl J.Stat.Phys.\ }}

\ref { Albrecht, A. (1992), to appear in, {\it Physical Origins of Time
Asymmetry}, eds. J.J.Halliwell, J.Perez-Mercader and W.H.Zurek (Cambridge
University Press, Cambridge). Two perspective on a decohering spin.}

\ref { Anderson, A. (1990), \pr {\bf D42}, 585. On predicting correlations from
Wigner functions.}

\ref { Banks, T. (1985), \np {\bf B249}, 332. TCP, quantum gravity, the
cosmological constant and all that...}

\ref { Banks, T., Fischler, W. and Susskind, L. (1985), \np {\bf B262}, 159.
Quantum cosmology in 2+1 and 3+1 dimensions.}

\ref { Barrow, J.D. and Tipler, F. (1986), {\it The Anthropic Cosmological
Principle} (Oxford University Press, Oxford).}

\ref { Barvinsky, A.O. and Kamenshchik, A.Yu. (1990), \cqg {\bf 7}, 2285.
Preferred basis in the many-worlds interpretation of quantum mechanics and
quantum cosmology.}

\ref { Bell, J.S. (1981), in {\it Quantum Gravity 2: A Second Oxford
Symposium}, eds. C.J.Isham, R.Penrose and D.W.Sciama (Clarendon Press, Oxford).
Quantum mechanics for cosmologists.}

\ref { Blencowe, M. (1991), \annp {\bf 211}, 87. The consistent histories
interpretation of quantum fields in curved spacetime.}

\ref { Calzetta, E. and Hu, B.L. (1989), \pr {\bf D40}, 380. Wigner
distribution and phase space formulation of quantum cosmology.}

\ref { Deutsch, D. (1985), \ijtp {\bf 24}, 1. Quantum theory as a universal
physical theory.}

\ref { DeWitt, B.S. (1967), \pr {\bf 160}, 1113. Quantum theory of gravity. I.
The canonical theory.}

\ref { DeWitt, B.S. and Graham, N. (eds.) (1973), {\it The Many Worlds
Interpretation of Quantum Mechanics} (Princeton University Press, Princeton).}

\ref { Dowker, H.F. and Halliwell, J.J. (1992), CTP preprint \#2071,
Fermilab-pub-92-44A, to appear in {\sl Phys.Rev.D}. The quantum mechanics of
history: the decoherence functional in quantum mechanics.}

\ref { Everett, H. (1957), \rmp {\bf 29}, 454. Relative state formulation of
quantum mechanics.}

\ref { Fang, L.Z. and Ruffini, R.(eds.) (1987), {\it Quantum Cosmology},
Advanced Series in Astrophysics and Cosmology No.3 (World Scientific,
Singapore).}

\ref { Fang, L.Z. and Wu, Z.C. (1986), \ijmp {\bf A1}, 887. An overview of
quantum cosmology.}

\ref { Farhi, E., Goldstone, J. and Gutmann, S. (1989), \annp {\bf 192}, 368.
How probability arises in quantum mechanics.}

\ref { Finkelstein, D. (1963), {\sl Trans.N.Y.Acad.Sci.} {\bf 25}, 621. The
logic of quantum physics.}

\ref { Fischler, W., Ratra, B. and Susskind, L. (1986), \np {\bf B259}, 730.
\rm ({\it errata} \np {\bf B268} 747 (1986)). Quantum mechanics of
inflation.}

\ref { Fukuyama, T. and Morikawa, M. (1989), \pr {\bf 39}, 462. Two-dimensional
quantum cosmology: directions of dynamical and thermodynamical arrows of time.}

\ref { Gell-Mann, M. and Hartle, J.B. (1990) in, {\it Complexity, Entropy
and the Physics of Information}, Santa Fe Institute Studies in the Sciences of
Complexity, vol IX, edited by W.H.Zurek (Addison Wesley); also in, {\it
Proceedings of the Third International Symposium on Foundations of Quantum
Mechanics in the Light of New Technology}, edited by S.Kobayashi (Japan
Physical Society). Quantum cosmology and quantum mechanics.}

\ref { Geroch, R. (1984), {\sl No{$\sl \hat u$}s} {\bf 18}, 617. The Everett
interpretation. (This issue of {\sl No{$\sl \hat u$}s} also contains a number
of other papers in which the Everett interpretation is discussed, largely from
a philosophical point of view.)}

\ref { Graham, N. (1973), in {\it The Many Worlds Interpretation of Quantum
Mechanics}, B.S.DeWitt and N.Graham (eds.) (Princeton University Press,
Princeton). The measurement of relative frequency.}

\ref { Griffiths, R.B. (1984), \jsp {\bf 36}, 219. Consistent histories and the
interpretation of quantum mechanics.}

\ref { Grishchuk, L.P. (1987), {\sl Mod.Phys.Lett.} {\bf A2}, 631. Quantum
creation of the universe can be observationally verified.}

\ref { Guth, A.H. (1981), \pr {\bf D28}, 347. The inflationary universe: a
possible solution to the horizon, flatness and monopole problems.}

\ref { Habib, S. (1990), \pr {\bf 42}, 2566. The classical limit in quantum
cosmology. I. Quantum mechanics and the Wigner function. }

\ref { Habib, S. and Laflamme, R. (1990), \pr {\bf 42}, 4056. Wigner function
and decoherence in quantum cosmology. }

\ref { Halliwell, J.J. (1987), \pr {\bf D36}, 3626. Correlations in the wave
function of the universe.}

\ref { Halliwell, J.J. (1989), \pr {\bf D39}, 2912. Decoherence in quantum
cosmology.}

\ref { Halliwell, J.J. (1990), in {\it Proceedings of the Seventh Jerusalem
Winter School for Theoretical Physics: Quantum Cosmology and Baby Universes},
eds. S.Coleman, J.B.Hartle, T.Piran and S.Weinberg (World Scientific,
Singapore). Introductory lectures on quantum cosmology.}

\ref { Halliwell, J.J. (1991), {\sl Scientific American} {\bf 265}, 76. Quantum
cosmology and the creation of the universe.}

\ref { Halliwell, J.J. (1992), CTP preprint \#2070, to appear in {\sl
Phys.Rev.D}. Smeared Wigner functions and quantum-mechanical histories.}

\ref { Halliwell, J.J. and Hartle, J.B. (1990), \pr {\bf D41}, 1815.
Integration contours for the no-boundary wave function of the universe.}

\ref { Halliwell, J.J. and Hartle, J.B. (1991), \pr {\bf D43}, 1170. Wave
functions constructed from an invariant sum-over-histories satisfy
constraints.}

\ref { Halliwell, J.J. and Hawking, S.W. (1985), \pr {\bf D31}, 1777. Origin of
structure in the universe.}

\ref { Hartle, J.B. (1968), \amjp {\bf 36}, 704. Quantum mechanics of
individual systems.}

\ref { Hartle, J.B. (1985), in {\it High Energy Physics 1985: Proceedings of
the Yale Summer School,} eds. M.J.Bowick and F.Gursey (World Scientific,
Singapore). quantum cosmology.}

\ref { Hartle, J.B. (1986), in {\it Gravitation in Astrophysics, Cargese,
1986}, eds. B.Carter and J.Hartle (Plenum, New York). Prediction and
observation in quantum cosmology.}

\ref { Hartle, J.B. (1990), in {\it Proceedings of the Seventh Jerusalem Winter
School for Theoretical Physics: Quantum Cosmology and Baby Universes}, eds.
S.Coleman, J.B.Hartle, T.Piran and S.Weinberg (World Scientific, Singapore).
The quantum mechanics of cosmology.}

\ref { Hartle, J.B. (1992a), this volume.}

\ref { Hartle, J.B. (1992b), to appear in {\it Proceedings of the
Les Houches Summer School, 1992}.}

\ref { Hartle, J.B. and Hawking, S.W. (1983), \pr {\bf D28}, 2960. Wave
function of the universe.}

\ref { Hawking, S.W. (1982), in {\it Astrophysical Cosmology}, eds.
H.A.Br\"uck, G.V.Coyne and M.S.Longair (Pontifica Academia Scientarium, Vatican
City) ({\sl Pont.Acad.Sci.Varia} {\bf 48}, 563). The boundary conditions of the
universe.}

\ref { Hawking, S.W. (1984a), \np {\bf B239}, 257. The quantum state of the
universe.}

\ref { Hawking, S.W. (1984b), in {\it Relativity, Groups and Topology II, Les
Houches Session XL}, eds. B.DeWitt and R.Stora (North Holland, Amsterdam).
Lectures on quantum cosmology.}

\ref { Hawking, S.W. and Page, D.N. (1986), \np {\bf B264}, 185. Operator
ordering and the flatness of the universe.}

\ref { Hawking, S.W. and Page, D.N. (1988), \np {\bf B298}, 789. How probable
is inflation?}

\ref { Hu, B.L. (1989), Cornell preprint, CLNS 89/941. Quantum and statistical
effects in superspace cosmology.}

\ref { Isham, C.J. (1991), Imperial College preprint TP/90-91/14,
to appear in {\it Proceedings of the Schladming Winter School, 1991}.
Conceptual and geometrical problems in quantum gravity.}

\ref { Joos, E. and Zeh, H.D. (1985), {\sl Zeit.Phys.} {\bf B59}, 223. The
emergence of classical properties through interaction with the environment.}

\ref { Kazama, Y. and Nakayama, R. (1985), \pr {\bf D32}, 2500. Wave packet in
quantum cosmology.}

\ref { Kent, A. (1990), \ijmp {\bf A5}, 1745. Against many-worlds
interpretations.}

\ref { Kiefer, C. (1987), \cqg {\bf 4}, 1369. Continuous measurement of
minisuperspace variables by higher multipoles.}

\ref { Kiefer, C. (1988), \pr {\bf D38}, 1761. Wave packets in minisuperspace.}

\ref { Kodama, H. (1988), Kyoto University preprint KUCP-0014. Quantum
cosmology in terms of the Wigner function. }

\ref { Kucha{$\rm \check r$, K.V. (1989)}, in {\it Proceedings of the Osgood
Hill Meeting on Conceptual Problems in Quantum Gravity}, eds. A.Ashtekar and
J.Stachel (Birkhauser, Boston). The problem of time in canonical quantization
of relativistic systems.}

\ref { Kucha\v r, K.V. (1992), in {\it Proceedings of the 4th Canadian
Conference on General Relativity and Relativistic Astrophysics}, eds.
G.Kunstatter, D.Vincent and J.Williams (World Scientific, Singapore, 1992).
Time and interpretations of quantum gravity.}

\ref { Kucha{$\rm \check r$, K.V. and Ryan, M.P.} (1986), in Yamada
Conference XIV, eds. H.Sato and T.Nakamura (World Scientific). Can
minisuperspace quantization be justified?}

\ref { Kucha{$\rm \check r$}, K.V. and Ryan, M.P. (1989), \pr {\bf D40}, 3982.
Is minisuperspace quantization valid? Taub in Mixmaster.}

\ref { Laflamme, R. (1991), in {\it Proceedings of the XXII GIFT International
Seminar on Theoretical Physics}, St. Feliu de Guixols, Spain, June 3--8, 1991.
Introduction and applications of quantum cosmology.}

\ref { Laflamme, R. and Louko, J. (1991), \pr {\bf D43}, 3317. Reduced density
matrices and decoherence in quantum cosmology.}

\ref { Lapchinsky, V. and Rubakov, V.A. (1979), {\sl Acta.Phys.Polonica} {\bf
B10}, 1041. Canonical quantization of gravity and quantum field theory in
curved space-time. }

\ref { Leggett, A.J. (1980), {\sl Suppl.Prog.Theor.Phys.} {\bf 69}, 80.
Macroscopic quantum systems and the quantum theory of measurement.}

\ref { Markov, M.A., and Mukhanov, V.F. (1988), \pl {\bf 127A}, 251. Classical
preferable basis in quantum mechanics.}

\ref { Mellor, F. (1989), \np {\bf B353}, 291. Decoherence in quantum
Kaluza-Klein theories.}

\ref { Misner, C.W. (1969), \pr {\bf 186}, 1319. Quantum cosmology I.}

\ref { Misner, C.W. (1972), in {\it Magic Without Magic: John Archibald
Wheeler, a Collection of Essays in Honor of his 60th Birthday}, ed. J.Klauder
(Freeman, San Francisco). Minisuperspace.}

\ref { Morikawa, M. (1989), \pr {\bf D40}, 4023. Evolution of the
cosmic density matrix.}

\ref { Omn\`es, R. (1990), \annp {\bf 201}, 354. From Hilbert space to common
sense: a synthesis of recent progress in the interpretation of quantum
mechanics.}

\ref { Omn\`es, R. (1992), \rmp {\bf 64}, 339. Consistent interpretations of
quantum mechanics.}

\ref { Padmanabhan, T. (1989), \pr {\bf D39}, 2924. Decoherence in the density
matrix describing quantum three-geometries and the emergence of classical
spacetime.}

\ref { Page, D.N. (1986), in {\it Quantum Concepts in Space and Time}, eds.
R.Penrose and C.J.Isham (Clarendon Press, Oxford). Hawking's wave function for
the universe.}

\ref { Page, D.N. (1991), in {\it Proceedings of the Banff Summer Institute on
Gravitation}, eds. R.B.Mann and P.Wesson (World Scientific, Singapore).
Lectures on quantum cosmology.}

\ref { Paz, J.P. (1991), \pr {\bf D44}, 1038. Decoherence and backreaction in
quantum cosmology: The origin of the semiclassical Einstein equations.}

\ref { Paz, J.P. and Sinha, S. (1991), \pr {\bf D45}, 2823. Decoherence and
backreaction in quantum cosmology: multidimensional minisuperspace examples.}

\ref { Shirai, I. and Wada, S. (1988), \np {\bf B303}, 728. Cosmological
perturbations and quantum fields in curved space-time.}

\ref { Singh, T.P. and Padmanabhan, T. (1989), \annp {\bf 196}, 296. Notes on
semiclassical gravity. }

\ref { Smolin, L. (1984), in {\it Quantum Theory of Gravity: Essays in Honour
of
the 60th Birthday of Bryce DeWitt} (Hilger, Bristol). On quantum gravity and
the many-worlds interpretation of quantum mechanics.}

\ref { Tipler, F. (1986), \prep {\bf 137}, 231. Interpreting the wave function
of the universe.}

\ref { Unruh, W.G. and Wald, R. (1989), \pr {\bf D40}, 2598. Time and the
interpretation of quantum gravity.}

\ref { Unruh, W.G. and Zurek, W.H. (1989), \pr {\bf D40}, 1071. Reduction of a
wavepacket in quantum Brownian motion.}

\ref { Vachaspati, T. and Vilenkin, A. (1988), \pr {\bf D37}, 898.
Uniqueness of the tunneling wave function of the universe.}

\ref { Vilenkin, A. (1989), \pr {\bf D39}, 1116. The interpretation of the wave
function of the universe.}

\ref { Wada, S. (1986), \np {\bf B276}, 729. Quantum cosmological perturbations
in pure gravity.}

\ref { Wada, S. (1988), {\sl Mod.Phys.Lett} {\bf A3}, 645. Interpretation and
predictability of quantum mechanics and quantum cosmology.}

\ref { Wheeler, J.A. (1963), in {\it Relativity, Groups and Topology}, eds.
C.DeWitt and B.DeWitt (Gordon and Breach, New York). Geometrodynamics
and the issue of the final state.}

\ref { Wheeler, J.A. (1968), in {\it Batelles Rencontres}, eds. C.DeWitt and
J.A.Wheeler (Benjamin, New York). Superspace and the nature of quantum
geometrodynamics.}

\ref { Wheeler, J.A. and Zurek, W.H. (1983), {\it Quantum Theory and
Measurement} (Princeton University Press, Princeton).}

\ref { Zurek, W.H. (1981), \pr {\bf D24}, 1516. Pointer basis of quantum
apparatus: Into what mixture does the wave packet collapse?}

\ref { Zurek, W.H. (1982), \pr {\bf D26}, 1862. Environment-induced
superselection rules.}

\ref { Zurek, W.H. (1992), to appear in, {\it Physical Origins of Time
Asymmetry}, eds. J.J.Halliwell, J.Perez-Mercader and W.H.Zurek (Cambridge
University Press, Cambridge). Preferred sets of states, predictability,
classicality and the environment-induced decoherence.}

\end